\documentclass[11pt,a4paper,DIV11]{scrartcl}
\pdfoutput=1
\usepackage{jheppub}
\usepackage[utf8]{inputenc}
\usepackage[T1]{fontenc}
\usepackage{lmodern}
\usepackage[british]{babel}
\usepackage{amsmath}
\usepackage{amssymb}
\usepackage{amsfonts}
\usepackage{float}
\usepackage{stmaryrd}
\usepackage{epsfig}
\usepackage{epstopdf}
\usepackage{graphicx,psfrag}
\usepackage{subfigure}
\usepackage{placeins,bbm}

\newcommand{\tr}{\ensuremath{\mathrm{tr}}}

\providecommand{\href}[2]{#2}

\newcommand{\erw}[1]{\langle #1 \rangle}
\newcommand{\Seff}{S_{\text{eff}}}
\newcommand{\Seffo}{S_{\text{eff}}^{(1)}}
\newcommand{\Sefft}{S_{\text{eff}}^{(2)}}
\newcommand{\ve}[1]{{\bf #1}}
\newcommand{\nt}{{N_\tau}}
\newcommand{\ir}{{\bf r}}
\newcommand{\ix}{{\bf x}}
\newcommand{\la}{\lambda}
\newcommand{\lat}{\tilde{\lambda}}

\title{Numerical corrections to the strong coupling effective Polyakov-line action for finite 
T Yang-Mills theory}
\author[a]{G.~Bergner}
\author{J.~Langelage}
\author[b]{O.~Philipsen}
\affiliation[a]{Universit\"at Bern, Institut f\"ur Theoretische Physik}
\affiliation[b]{Institut f\"ur Theoretische Physik,
Goethe-Universit\"at Frankfurt, \\ Max-von-Laue-Str. 1, 60438 Frankfurt am Main, Germany}

\emailAdd{bergner@itp.unibe.ch}
\emailAdd{katzeee@gmx.de}
\emailAdd{philipsen@th.physik.uni-frankfurt.de}

\abstract{
We consider a three-dimensional effective theory of Polyakov lines derived previously from lattice 
Yang-Mills theory and QCD by means of a resummed strong coupling expansion. 
The effective theory is useful for investigations of the phase structure,
with a sign problem mild enough to allow simulations also at finite density.
In this work we present a numerical method to determine improved values for the effective couplings
directly from correlators of 4d Yang-Mills theory. 
For values of the gauge coupling up to the vicinity of the phase transition, 
the dominant short range effective coupling are well described by their
corresponding strong coupling series. 
We provide numerical results also for the longer range interactions, 
Polyakov lines in higher representations as well as four-point interactions, and
discuss the growing significance of non-local contributions as 
the lattice gets finer.
Within this approach the critical Yang-Mills coupling $\beta_c$ is reproduced to better than
one percent from a one-coupling effective theory on $N_\tau=4$ lattices while up to five
couplings are needed on $N_\tau=8$ for the same accuracy.
}

\begin{document}
\maketitle

%%%%%%%%%%%%%%%%%%%%%%%%%%%%%%%%%%%%%%%%%%%%%%%%%%%%%%%%%%%%%%%%%%%%%%%%
\section{Introduction}

Yang-Mills theory as well as QCD can be mapped to three-dimensional effective 
Polyakov loop theories by integrating over the spatial gluon degrees of freedom, both in the
continuum \cite{models,vuorinen,janp,plhq} and on the lattice 
\cite{Svetitsky:1982gs,Wozar:2007tz,Langelage:2010yr,Langelage:2014vpa,green1,green2}. 
The resulting effective theories 
describe the physics of the phase transitions and thermodynamic behaviour.
They are simpler in the sense that their non-perturbative treatment requires less numerical 
effort and, at least in some limiting cases, the effective interactions can be 
associated with the low energy degrees of freedom. 
Since part of the degrees of freedom have been integrated out 
the sign problem, which hampers lattice 
simulations at finite baryon density, is much milder in the effective theory. 
Indeed, at least for very heavy quarks \cite{Langelage:2014vpa} 
and in the chiral limit on very coarse lattices \cite{deForcrand:2014tha} analytically derived 
effective lattice theories allow for
simulations of the nuclear liquid gas transition in the cold and dense regime of QCD.
%There is also evidence that perturbative treatments of the effective theory 
%provide reasonable results \cite{}.
On the other hand, effective 
actions generically include infinitely many and arbitrarily involved interaction terms, 
restricted only by the symmetries of the underlying theory, and 
become practical only after some drastic truncation. It is therefore essential to understand the 
relevance of the various terms and the effects of truncations.

In this work we consider Yang-Mills theory on a lattice with periodic boundary conditions 
and the effective action obtained
by integrating over the spatial links,
\begin{equation}
e^{-\Seff[W]}=\int [dU_i] e^{-S[U]}\; ,
 \label{eq:basicaction}
\end{equation}
where $S[U]$ is the standard Wilson lattice gauge action and
$\Seff$ depends only on the Polyakov lines $W(\ve{x})=\prod_{y_0;\; \ve{y}=\ve{x}} U_0(y)$.
It contains all $n$-point interactions at all distances and with all powers of loops that are
consistent with the $Z(N)$-centre symmetry of the Yang-Mills theory.  
The expectation value of all observables represented by Polyakov lines can 
be calculated in the effective theory,
\begin{equation}
\erw{O}=\frac{1}{Z}\int [dW] e^{-\Seff[W]} O(W)=\frac{1}{Z}\int [dU_0][dU_i] e^{-S[U]} O(W[U])\; .
 \label{eq:integration}
\end{equation}
The normalisation of the path integral is chosen such that
\begin{equation}
Z= \int [dU_i] e^{-S[U]}=\int [dW] e^{-\Seff[W]}\; .
\end{equation}
The constant part of the effective action, which does not depend on the $W$, can be 
neglected (absorbed in the normalisation) if one is interested in the phase transition or $W$-dependent observables only. However, for detailed thermodynamic properties the $T$ dependence 
of the constant part might be relevant.

Svetitsky and Yaffe conjectured that the short range Polyakov interactions are the relevant terms 
for the phase transition \cite{Svetitsky:1982gs}. This is based on the fact that these are the dominant contributions both in the strong and weak coupling limit, while for 
all couplings interactions are screened by the mass gap of the theory (see also the discussion
in \cite{Bergner:2013qaa}).
In \cite{Langelage:2010yr} the integration of the spatial links was performed by means of a strong coupling expansion, where interaction terms are parametrically suppressed with
increasing distance and $n$-points. 
Simulation of the effective action truncated to the leading nearest neighbour coupling gives
already a good description of thermodynamic functions and  the phase transition, allowing for
a continuum extrapolation of the critical temperature. This already vastly improves over 
direct strong coupling calculations of thermodynamic observables \cite{scthermo}.
On the other hand, the missing higher order couplings and long range interactions
hamper the applicability of the effective theory to correlation functions and the string tension, where non-local contributions play an increasing role as lattices get 
finer \cite{Bergner:2013qaa}. The same observation was made with a non-perturbatively
determined form of the effective action \cite{green1,green2}.

In this work we present
a new method to numerically determine the couplings of the effective theory directly 
from simulations of the 4d Yang-Mills theory. The method employs the fact that there are different 
character expansions of the effective action. In one of them the effective couplings emerge
as expectation values of $n$-point functions of Polyakov loops in the full theory,
which can be measured easily. The relation between effective couplings corresponding
to different character expansion schemes is then established perturbatively
in terms of a rapidly converging power series.
This allows us to check the range of applicability of the strong coupling approach 
\cite{Langelage:2010yr,Bergner:2013qaa} and 
significantly improve on the results.

This paper is organised as follows. In section \ref{sec:character}  we discuss the character
expansions pertaining to the strong coupling approach and the new numerical method.
We investigate two-point interactions of Polyakov 
loops in the fundamental and anti-fundamental representation in section \ref{sec:fundtwo} . 
The new method is introduced for the one-coupling effective theory and then
generalised to include two-point couplings over larger distances.
In Section \ref{sec:hirep} we discuss the significance of higher representations as well as three- and four-point interactions in the fundamental representation before concluding in 
section \ref{sec:conc}.

\section{The different character expansions of the effective action}
\label{sec:character}

The effective action can be expanded in terms of the characters  $\chi_\ir$ of all irreducible representations of the gauge group. 
The irreducible representations of SU(N) are labelled by a vector $\ir$ of N-1 canonical labels.
In SU(3) one has two-dimensional vectors like $(1,0)$ and $(0,1)$ for fundamental and anti-fundamental representation or
$(1,1)$ and $(2,0)$ for the adjoint and sextet representation.
The characters of the fundamental and anti-fundamental representation are the ordinary Polyakov loop $L(\ve{x})\equiv\chi_{(1,0)}(W(\ve{x}))=\tr[W(\ve{x})]$ and its complex conjugate $L^\dag\equiv\chi_{\overline{(1,0)}}(W)=\chi_{(0,1)}(W)$.
Since products of the characters of the irreducible representations can be transformed into sums, the effective action can be transformed 
into a representation without products of $\chi_\ir(W(\ix))$ at the same lattice point $\ix$.
The effective action generically includes arbitrary interactions of characters at all distances. 
A possible representation of the effective action that includes all of these terms is thus
\begin{equation}
 \Seff=\sum_{\ix,\ir}\sum_n \sum_{\begin{array}{c} \scriptscriptstyle [\ir_1,\ix_1],\ldots, [\ir_n,\ix_n]
 %;\\ \scriptscriptstyle |\ix|\leq|\ix_1|\leq|\ix_2|\ldots 
 \end{array} } 
 c^\ir_{
[\ir_1,\ix_1],\ldots,[\ir_n,\ix_n]}\chi_{\ir}(W(\ix))\prod_{i=1}^n \chi_{\ir_i}(W(\ix+\ix_i))\; .
 \label{eq:Seffexp}
\end{equation}
The coefficients of this expansion are not completely independent but related by the symmetries of the theory. In our case, the $Z(N)$ centre symmetry restricts the independent combinations of the $\ir_i$, and the cubic symmetry 
(the remnant of rotational and translational invariance in the continuum) 
restricts the combinations of $\ix_i$.

Our aim is a numerical determination of the coefficients in the effective action from expectation values in the full theory. 
The relations between the measured expectation values and the effective action should be as simple as possible.
Certain observables like the Polyakov loop correlators can be obtained with a high precision.
The form \eqref{eq:Seffexp} is, however, not well suited for such an approach.

Alternatively, also the exponential of the effective action can be represented by
interaction terms of characters.
This corresponds to a Taylor expansion of the right hand side of \eqref{eq:basicaction}, where products of characters
at the same point are converted into sums of single characters. 
In this way the expansion of the effective action \eqref{eq:Seffexp} is converted into an expansion of its exponential,
\begin{equation}
  e^{-\Seff}=
\tilde\lambda_0 \bigg(1+
\sum_{\ix,\ir}\sum_n \tilde{\sum_{\begin{array}{c} \scriptscriptstyle [\ir_1,\ix_1],\\ \scriptscriptstyle \ldots, [\ir_n,\ix_n] \end{array} }}
\tilde{\lambda}^\ir_{[\ix_1,\ir_1],\ldots,[\ix_n,\ir_n]}\chi_{\ir}(W(\ix))\prod_{i=1}^n \chi_{\ir_i}(W(\ix+\ix_i))\bigg)\; .
  \label{eq:expSeffexpsum}
\end{equation}
The sum $\tilde{\sum}$ is performed in such a way that each combination of characters and lattice points appears only once. Because of the orthogonality of the characters,
expectation values of the characters in the full theory now directly 
project on the corresponding coupling constant,
\begin{eqnarray}
\label{eq:expSeffexpsumproj}
\langle \chi_{\ir}(W(0))\prod_{i=1}^n \chi_{\ir_i}(W(\ix_i))\rangle
&=&\frac{1}{Z} \int [dU_\mu] e^{-S[U]}\chi_{\ir}(W(0))\prod_{i=1}^n \chi_{\ir_i}(W(\ix_i))\\
&=&\frac{1}{Z} \int [dW] e^{-\Seff}\chi_{\ir}(W(0))\prod_{i=1}^n \chi_{\ir_i}(W(\ix_i))=\tilde\lambda^\ir_{[\ix_1,\ir_1],\ldots,[\ix_n,\ir_n]}\; . \nonumber
\end{eqnarray}
(The constant part $\tilde\lambda_0=\int [dU_\mu] e^{-S[U]}$ has been factored out to arrive at the proper normalisation of the expectation values and
will not be considered in the further analysis).
The disadvantage of this representation of the effective action is, however, that any truncated version of the expansion is not appropriate to reproduce the phase transition. 
The range of interactions in the corresponding effective theory is restricted to short ranges. At larger distances all correlations, in particular the Polyakov loop correlator $\erw{L(\ve{0}) 
L^\dag(\ix)}$, are zero.

A different representation, which includes all long range correlations already in the truncation to its 
first term,
follows in a natural way from the strong coupling approach 
\cite{Langelage:2010yr},
\begin{align}
  e^{-\Seff}=\lambda_0 
 \prod_{\ix,\ir,n} \tilde{\prod_{\begin{array}{c} \scriptscriptstyle [\ir_1,\ix_1],\\ \scriptscriptstyle \ldots, [\ir_n,\ix_n] 
\end{array} }}
  \bigg(1+
\lambda^\ir_{[\ix_1,\ir_1],\ldots,[\ix_n,\ir_n]}\bigg(\chi_{\ir}(W(\ix))\prod_{i=1}^n \chi_{\ir_i}(W(\ix+\ix_i))+c.c.\bigg) \bigg)\; .
  \label{eq:expSeffexp}
\end{align}
The product $\tilde{\prod}$ now runs over all inequivalent combinations of irreducible representations $\ir$ and lattice positions $\ix$. The coefficient of the trivial representation has been factored out since it is irrelevant in the calculation of expectation
values. A truncation in this representation is a restriction of $n$ and $\ir,\ir_1,\ldots, \ir_n$  to a certain number of
irreducible representations and of the $\ix_i$ to a finite number of interaction distances. 
Such truncations arise naturally in a strong coupling expansion. 
This representation combines the advantages of \eqref{eq:Seffexp} and \eqref{eq:expSeffexpsumproj}: 
long range correlations are included and the calculation of the coefficients from expectation values is still feasible.
The Polyakov loop correlator in such an effective theory is non-zero at arbitrary
distances and decays exponentially, as in the
full gauge theory. 

All discussed representations are complete expansions of the effective action, implying relations between the coefficients $c$, $\lambda$, and $\tilde\lambda$. If the expansions are truncated, only approximate relations can be established. In this work we use the representation (\ref{eq:expSeffexpsumproj}) in order to determine a finite number of the couplings $\tilde{\lambda}$
numerically by simulating the corresponding $n$-point functions. In a second step, we evaluate
the same expectation values as a perturbative series in small $\lambda$ using the representation
(\ref{eq:expSeffexp}). 
The relations $\tilde\lambda(\lambda)$ are then realised as power series which
can be easily solved for the $\lambda$. 
In this approach the effective couplings, containing information of the full correlators, 
clearly deviate from those obtained in the 
strong coupling expansion. But we find them to be
sufficiently small over the entire parameter range of $\beta,\nt$ to
justify a perturbative expansion, with an
ordering essentially the same as in the strong coupling expansion. In particular, 
the dominant contribution is the nearest neighbour interaction.

Above the phase transition strong coupling series for the effective couplings are no longer valid, whereas their numerical
determination is of course still possible. However, the effective action now
has to take into account the non-zero expectation value of the fundamental loop (as well as that of Polyakov loops in higher representations with non-zero N-ality). One possibility is to add a one-point term, here we include the expectation value by a simple shift of the Polyakov loop variables,
\begin{equation}
 L\rightarrow L-\erw{L}\; .
 \label{eq::shift}
\end{equation}
It is important to note that the effective theory \eqref{eq:expSeffexp} captures the signal of the phase transition even without this shift, the non-vanishing expectation value resulting from the
non-perturbative dynamics. By contrast, this is impossible in the representation \eqref{eq:expSeffexpsum} without addition of a one-point term.
Our main interest is in the effective theory up to the phase transition and its prediction 
for the transition point. For a better presentation of the effective couplings we include in 
some cases data above the transition, including the shift \eqref{eq::shift} .

%%%%%%%%%%%%%%%%%%%%%%%%%%%%%%%%%%%%%%%%%%%%%%%%%%%%%%%%%%%%%%%%%%%%%%%%
\section{Two point interactions in the fundamental representation}
\label{sec:fundtwo}

The simplest contributions to the effective action are two-point interaction terms of Polyakov loops in the fundamental and anti-fundamental representation at all distances.
We first list the strong coupling results and the corresponding ordering of the couplings, then we compare it with the non-perturbative numerical data.

\subsection{Strong coupling results}
The expansion parameter of the strong coupling series is $u(\beta)=\frac{\beta}{18}+\ldots < 1$, the coefficient of the fundamental representation in a character expansion 
of $e^{-S_p[U]}$ with the plaquette action $S_p(U)$ \cite{Montvay:1994cy}. 
To leading order 
the nearest neighbour coupling is $u^{\nt}$, where $\nt$ is the number of lattice points in temporal direction. 
The leading order contribution to couplings over
larger distances is at least $d\nt$, where $d$ is the shortest distance along the links of the lattice (``taxi driver distance'').
We write the effective couplings always as a product of leading order contribution and corrections.
For the nearest neighbour interaction we have computed the corrections up to high orders in $u$,
they can be resummed into an exponential factor (for details see \cite{Langelage:2010yr}),
\begin{align}
 &\lambda_1\equiv \lambda^{(1,0)}_{[(1,0,0),(0,1)]} =\nonumber \\
& \left\{ 
\begin{aligned}
u^4 \exp\left[ 4 (4 u^4+12 u^5-14 u^6-36 u^7+\frac{295}{2} u^8+\frac{1851}{10} u^9+\frac{1035317}{5120} u^{10}) \right]&:\; \nt = 4\\
u^{\nt} \exp\left[\nt (4u^4+12u^5-14u^6-36u^7+\frac{295}{2} u^8+\frac{1851}{10}u^9+\frac{1055797}{5120}u^{10})\right] &:\; \nt \geq 6 \; .
 \end{aligned}\right.
 \label{eq:l1sc}
\end{align}
Here we introduced a simplified one-index notation for the couplings $\lambda$, which we also apply for the representation \eqref{eq:expSeffexpsum} of the effective action ($\tilde\lambda_1\equiv \tilde\lambda^{(1,0)}_{[(1,0,0),(0,1)]}$ etc.).

The next terms are interactions at distance $\sqrt{2}$ and $2$, which have the same ``taxi-driver distance'' $d=2$:
\begin{align}
 &\lambda_2\equiv\lambda^{(1,0)}_{[ (1,1,0),(0,1)]}=
 \left\{
 \begin{aligned}
  u^{10} (12+26 u^2+364 u^4)&: \;\nt=4\\
 u^{14} (30 +66 u^2)&: \;\nt=6 \\
  \nt (\nt-1)u^{ 2\nt+2 } &: \; \nt > 6
 \end{aligned}\right.\\
 &\lambda_3\equiv \lambda^{(1,0)}_{[ (2,0,0),(0,1)]}= 4\nt u^{2\nt+6}(1 + 12u^2 + (8\nt+\frac{57}{2})u^4)\;.
\end{align}
We also consider interactions with $d=3$
\begin{align}
&\lambda_4\equiv \lambda^{(1,0)}_{[(1,1,1),(0,1)]}= u^{3\nt+4}(\nt^4+6\nt^3-13\nt^2+6\nt)\\
&\lambda_5\equiv\lambda^{(1,0)}_{[(2,1,0),(0,1)]}= \nt(\nt-1)u^{3\nt+4}\\
& \lambda_6\equiv \lambda^{(1,0)}_{[ (3,0,0),(0,1)]}= 4\nt u^{3\nt+8}(1 + 12u^2)\;.
\end{align}
Note that the corrections to the leading order strong coupling result have been computed with different precision for the various couplings since their calculation gets much more involved at larger distances.
In the following we limit our investigations to a maximum order of $4\nt$ on the $\nt=4$ lattice,
which excludes the interaction $\lambda_6$.
The effective action \eqref{eq:expSeffexp} has thus the following form 
\begin{eqnarray}
 e^{-\Seffo}&=&\prod_{x,i=1,\ldots 3} \Big(1+\lambda_1(L(x) L(x+\hat i)^\dag+L(x)^\dag L(x+\hat i))\Big)\nonumber\\
 && \prod_{[x,y]} \Big(1+\lambda_2(L(x) L(y)^\dag+L(x)^\dag L(y))\Big)\nonumber\\
&& \prod_{x,i=1,\ldots 3} \Big(1+\lambda_3(L(x) L(x+2\hat i)^\dag+L(x)^\dag L(x+2\hat i))\Big)\nonumber\\
 && \prod_{\{x,y\}} \Big(1+\lambda_4(L(x) L(y)^\dag+L(x)^\dag L(y))\Big)\nonumber\\
 && \prod_{<x,y>} \Big(1+\lambda_5(L(x) L(y)^\dag+L(x)^\dag L(y))\Big)\; ,
 \label{eq:fundaction}
\end{eqnarray}
where  $[x,y]$ indicates the product of all points with distance $\sqrt{2}$ (i.~e.\ connected around one corner), the product $\{x,y\}$ contains all points with distance $\sqrt{3}$, and $<x,y>$ with distance $\sqrt{5}$.
%%%%%%%%%%%%%%%%%%%%%%%%%%%%%%%%%%%%%%%%%%%%%%%%%%%%%%%%%%%%%%%%%%%%%%%%%
\subsection{The effective coupling in the one-coupling action}

\begin{figure}[ht]
 \begin{center}
\includegraphics[width=14cm]{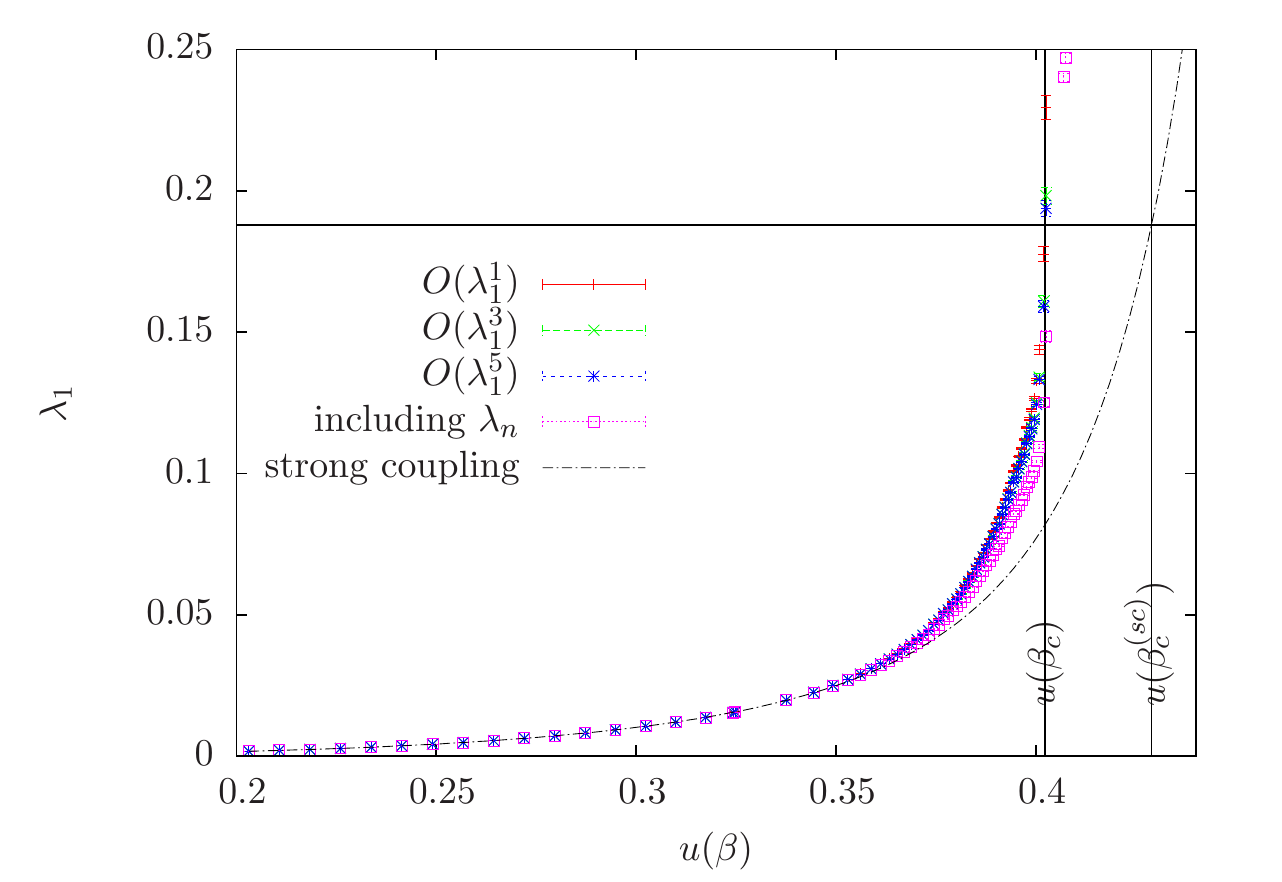}
 \end{center}
 \caption[]{Numerical solution for $\lambda_1$ in the one-coupling effective theory compared
 to its strong coupling series \eqref{eq:l1sc} for $\nt=4$. 
 Different truncations of \eqref{eq:npl1} and inclusion of some larger distance couplings 
 \eqref{eq:npl1l2l3t} (including $\lambda_n$) are  shown.
 The horizontal line marks the critical effective coupling,
 vertical lines the critical $\beta_c$ in full Yang-Mills theory and
 $\beta_c^{(sc)}$ predicted by $\lambda_1^{(sc)}$ \cite{Langelage:2010yr}.  
}
 \label{fig:smalll}
\end{figure}
The simplest truncation of the effective action contains only the nearest neighbour coupling $\lambda_1$, setting all other
couplings in \eqref{eq:fundaction} to zero. Previous investigations based on the strong coupling series
(\ref{eq:l1sc}) have shown
that this simple form of the action allows already a description of the phase 
transition to an accuracy of about 10\% for lattices with $N_\tau=2-12$ \cite{Langelage:2010yr}.

On the other hand, the exact numerical value of the coupling in the expansion \eqref{eq:expSeffexpsum}  
is determined by the simple correlator, 
\begin{equation}
 \tilde\lambda_1=\erw{L((0,0,0))L((1,0,0))^\dag}\; ,
\end{equation}
which we measure in a simulation of the full 4d theory.
The same correlator can be calculated in the effective theory up to $O(\lambda_1^7)$ in 
an expansion of $\lambda_1$.
Equating these two expressions one arrives at the following relation between the coupling constants in representation \eqref{eq:expSeffexpsum} and \eqref{eq:expSeffexp},
\begin{equation}
\tilde\lambda_1= \lambda_1+4\lambda_1^3+24 \lambda_1^5+O(\lambda_1^7)\; ,
 \label{eq:npl1}
\end{equation}
which can be solved numerically for $\lambda_1$. Note that $\tilde\lambda_1$ is then
exact (up to statistical errors and finite volume corrections), whereas a systematic error
is introduced to $\lambda_1$ by the truncation of the right hand side.

Figure \ref{fig:smalll} shows the results obtained for an $N_\tau=4$ lattice.
The strong coupling series gives a reasonable approximation to $\lambda_1$ in a wide 
parameter range. We observe that our numerical approach is self-consistent: 
even though $\lambda_1$ deviates from the strong coupling expression \eqref{eq:l1sc} at larger 
$u$, its value remains small and the truncation of  
\eqref{eq:npl1} is justified. Orders beyond $O(\lambda_1^3)$ in \eqref{eq:npl1} 
are not relevant and the influence of the interactions at larger distances ($\lambda_n$) appears 
to be suppressed. Details of the computation including these interactions are 
explained in the next section.

The phase transition in the one-coupling effective theory is at $(\lambda_1)_c=0.1879$ 
\cite{Langelage:2010yr}, i.e.~the intersection of the curve $\lambda_1(\beta)$ with this value
gives the prediction of the one-coupling theory for $\beta_c$.
Note that for the numerically determined coupling $\lambda_1$ this coincides with the value at the
true $\beta_c$ of full Yang-Mills,
$\lambda_1(\beta_c)\approx(\lambda_1)_c$. Thus, 
the one-coupling model with the effective coupling determined by our numerical approach provides a
quantitatively correct prediction for the transition point at $\nt=4$ and significantly improves over
the strong coupling result, which tends to overestimate $\beta_c$. However, we shall now see that
for larger $N_\tau$, i.e.~on finer lattices, more couplings are necessary for the same precision.

%%%%%%%%%%%%%%%%%%%%%%%%%%%%%%%%%%%%%%%%%%%%%%%%
\subsection{Interaction terms at larger distances}

\begin{figure}
 \begin{center}
 \subfigure[\label{fig:longrangeca}Measured correlations  ($\nt=4$)]{\includegraphics[width=7cm]{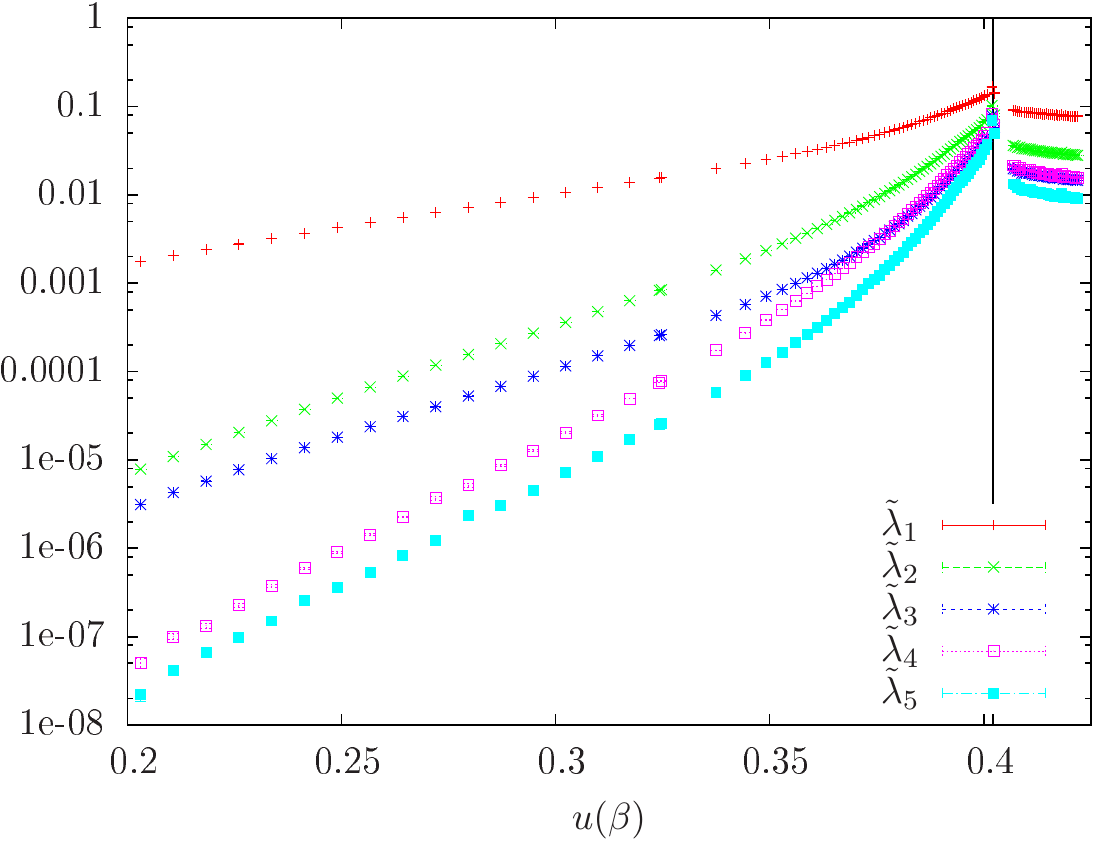}}
 \subfigure[\label{fig:longrangecb}Effective couplings ($\nt=4$)]{\includegraphics[width=7cm]{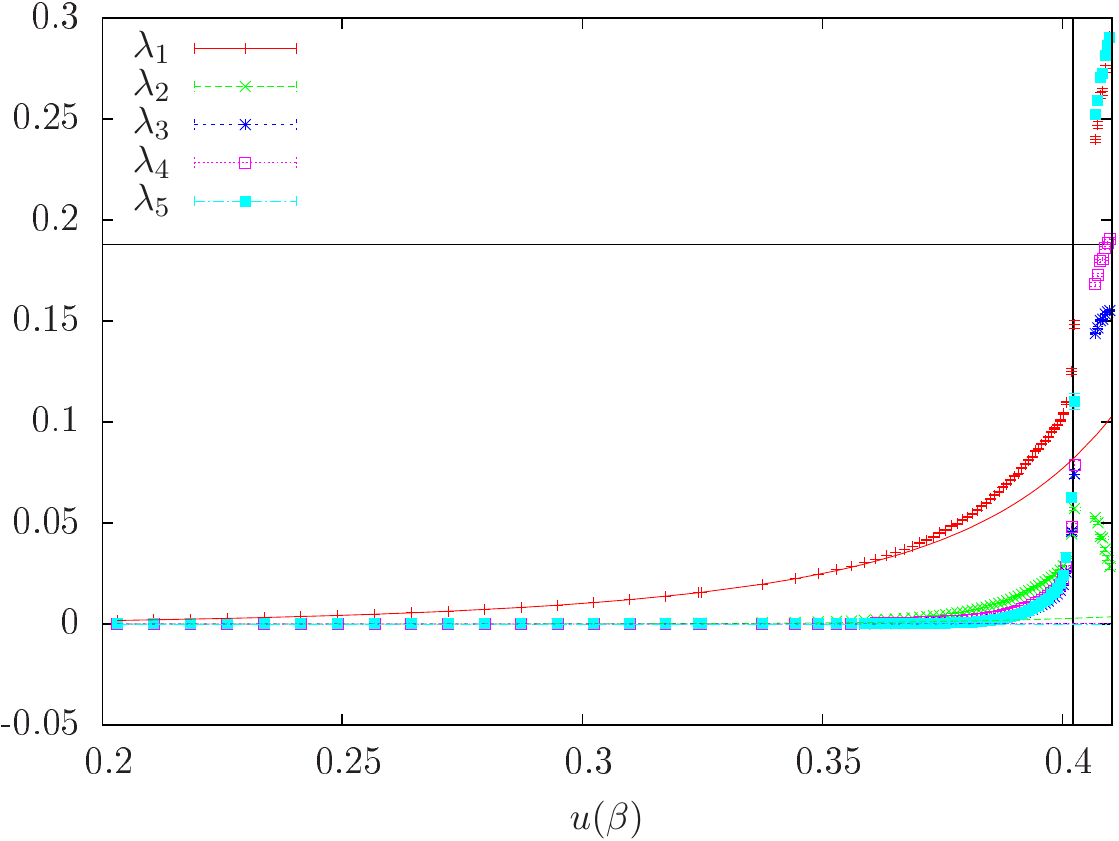}}
 \subfigure[\label{fig:longrangecc}Logarithmic representation ($\nt=4$)]{\includegraphics[width=7cm]{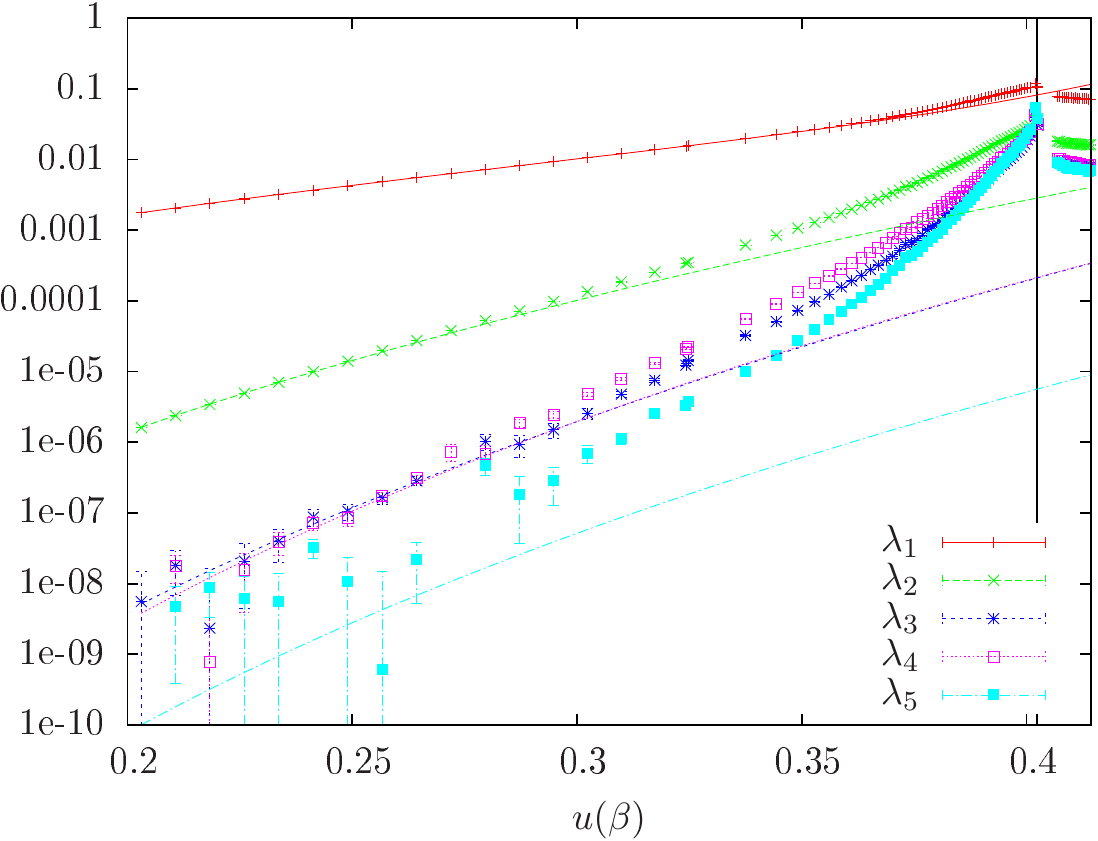}}
  \subfigure[\label{fig:longrangecd}Effective couplings ($\nt=6$)]{\includegraphics[width=7cm]{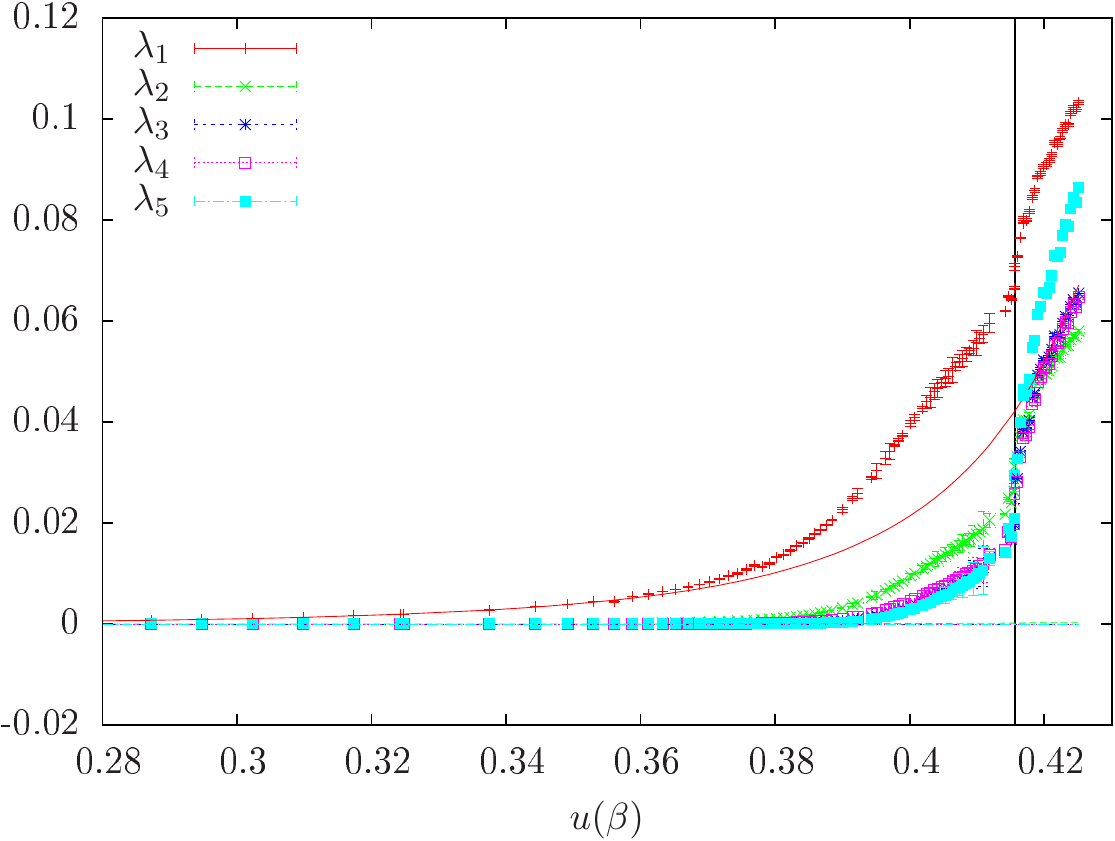}}
 \end{center}
 \caption{ (a) Effective couplings defined by correlators, \eqref{eq:expSeffexpsumproj}, from 
 simulations of 4d Yang-Mills on $\nt=4$.  
 (b) Solution of equations \eqref{eq:npl1l2l3t}. Lines show the strong coupling series for the couplings.
 (c) Same as (b) in logarithmic representation.
 (d) Same as (b) for $\nt=6$.
Vertical lines mark the critical coupling $u(\beta_c)$ in full Yang-Mills theory.
}
 \label{fig:longrangec}
\end{figure}

We now include all two-point interactions of fundamental and anti-fundamental loops whose strong
coupling series contain contributions up to $O(u^{4\nt})$, \eqref{eq:fundaction}.
Again we directly determine 
the couplings $\lat_n$ in representation \eqref{eq:expSeffexpsum}  by the corresponding 
Polyakov loop correlators,
\begin{equation}
 \tilde\lambda_2=\erw{L((0,0,0))L((1,1,0))^\dag}\; ;\quad  \tilde\lambda_3=\erw{L((0,0,0))L((2,0,0))^\dag}\; \quad \ldots\; .
\end{equation}
As in the one-coupling case, the same correlators are computed in the effective theory \eqref{eq:fundaction} using
 an expansion in the effective coupling constants $\lambda_n$. 
However, in this case the truncation is less obvious. With infinitely many couplings 
there are also infinitely many mixed terms with the same overall number of $\lambda$'s, i.e.~there
is no simple power counting.
We therefore use again the strong coupling expansion
as a guiding principle and truncate terms whose
leading order contribution would exceed $O(u^{4\nt})$.  
In that way a coupled system of equations is obtained, 
\begin{align}
\begin{split}
 \lat_1 &= \la_1 + 4 \la_1^3+8\la_1\la_2 +\ldots\\
 \lat_2 &=\la_2+2\la_1^2+4\la_2^2+16 \la_1^4+36 \la_1^2 \la_2+\ldots\\
 \lat_3 &= \la_3 + \la_1^2+4\la_2^2+24\la_1^2\la_2+12 \la_1^4 +\ldots\\
 \lat_4 &=\la_4+6\la_1^3+6\la_1\la_2+\ldots\\
 \lat_5 &=\la_5+3\la_1^3+2\la_1\la_2+\ldots\; ,
\end{split}
 \label{eq:npl1l2l3t}
\end{align}
that can be solved for the couplings $\la_n$. To estimate the effects of the truncation in the expansion of the couplings
we have also considered a second set of equation,
\begin{align}
\begin{split}
 \lat_1 &= \la_1 + 4 \la_1^3+8\la_1\la_2 +\ldots\\
 \lat_2-2\lat_1^2 &=\la_2+4\la_2^2+4 \la_1^2 \la_2 +\ldots\\
 \lat_3-\lat_1^2 &= \la_3 +4\la_2^2+8\la_1^2\la_2+4 \la_1^4 +\ldots\\
 \lat_4-6\lat_1^3 &=\la_4+6\la_1\la_2+\ldots\\
 \lat_5-3\lat_1^3 &=\la_5+2\la_1\la_2+\ldots\; ,
\end{split}
 \label{eq:npl1l2l3t2}
\end{align}
where the complete nearest neighbour correlation is subtracted. This corresponds to a resummation of certain dominant higher order terms. We get consistent results for both sets of equations, which is an indication that the truncation errors of the system of equations is small enough. In the following we will
use \eqref{eq:npl1l2l3t} for the determination of the $\la_n$. 
Figure \eqref{fig:longrangeca} shows the corresponding results and 
illustrates the accuracy of the Polyakov loop correlators required in order to
compare with the the strong coupling predictions.
\begin{figure}[ht]
 \begin{center}
\includegraphics[width=14cm]{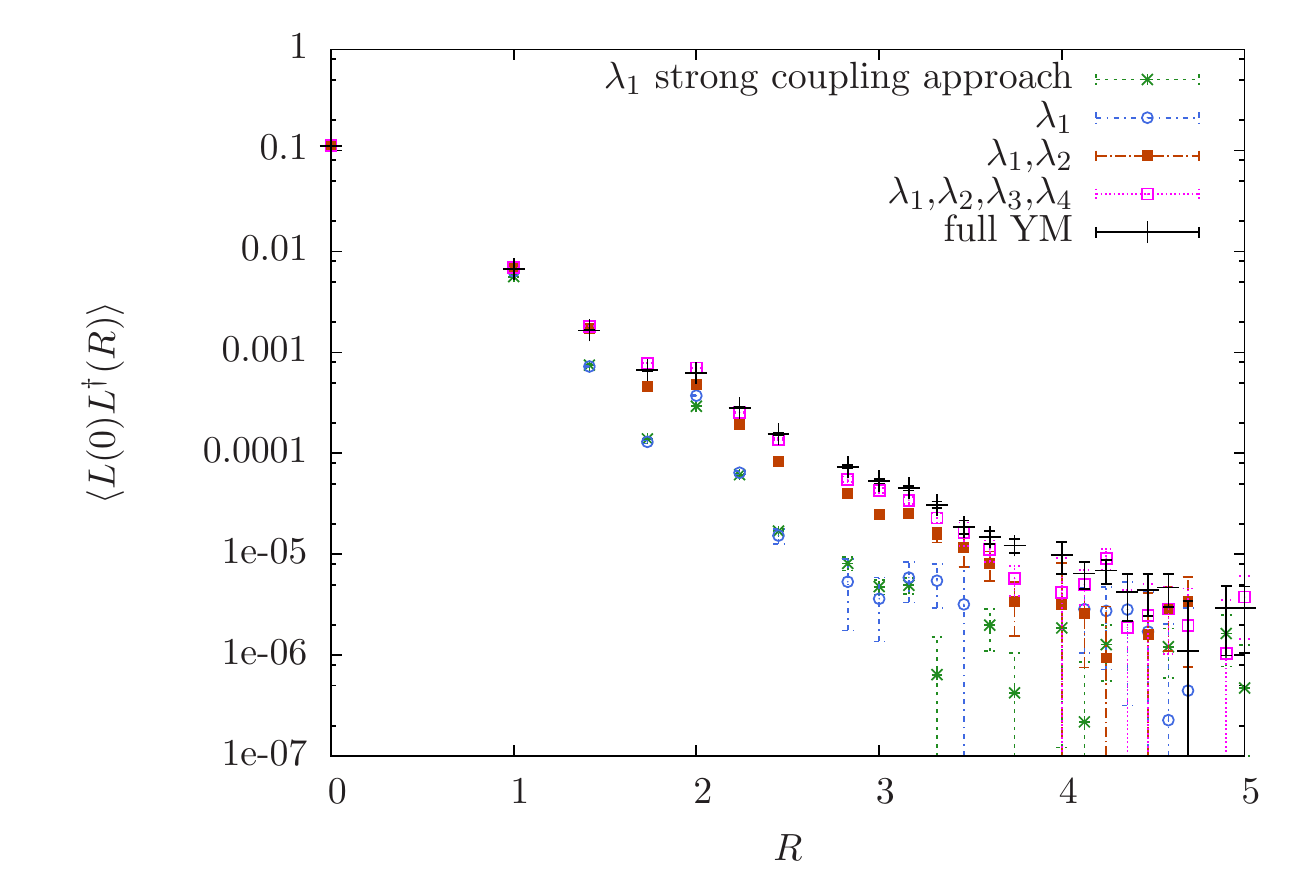}
 \end{center}
 \caption{Polyakov loop correlator in full Yang-Mills and the effective theory \eqref{eq:fundaction} 
 at $\beta=5.4$ on a $24^3\times 4$ lattice. The different truncations of the effective 
 theories  are based on the couplings
 determined by \eqref{eq:npl1l2l3t}, as well as the strong coupling series \eqref{eq:npl1} 
 for the one-coupling theory.}
 \label{fig:corrcomp}
\end{figure}
The numerical solution for the set of equations \eqref{eq:npl1l2l3t} is shown in Figure \ref{fig:longrangecb} and \ref{fig:longrangecc}. At small $\beta$ the result agrees with the prediction from the strong coupling expansion, which provides a good approximation for the couplings $\lambda_1$ and 
$\lambda_2$ over a wide range. The disagreement between the numerical determination and the strong coupling expansion is larger for the long range interactions and presumably due to the much shorter series for these couplings.

While the phase transition on an $\nt=4$ lattice is predicted with a good accuracy in the one-coupling effective theory, the long range interactions are necessary to reproduce the Polyakov line correlator of the 4d Yang-Mills theory. 
This is illustrated in Figure \ref{fig:corrcomp}. In \cite{Bergner:2013qaa} we observed 
that the mis-match of the correlator remains when additional couplings are
added in the strong coupling approach, since their leading order contributions are too small. The 
situation is different with the numerically determined effective couplings. While the correlator in the 
one-coupling theory still shows large deviations from the full one,
inclusion of two-point interactions up to $\la_4$ results in good agreement.  Thus the main improvement 
achieved by our numerical procedure are better estimates 
of the longer range interactions, where no sufficiently long strong coupling series are available.

%%%%%%%%%%%%%%%%%%%%%%%%%%%%%%%%%%%%%%%%%%
\subsection{Different $\nt$ and the continuum limit}
\label{sec:contlim}

\begin{figure}[t]
 \begin{center}
\includegraphics[width=12cm]{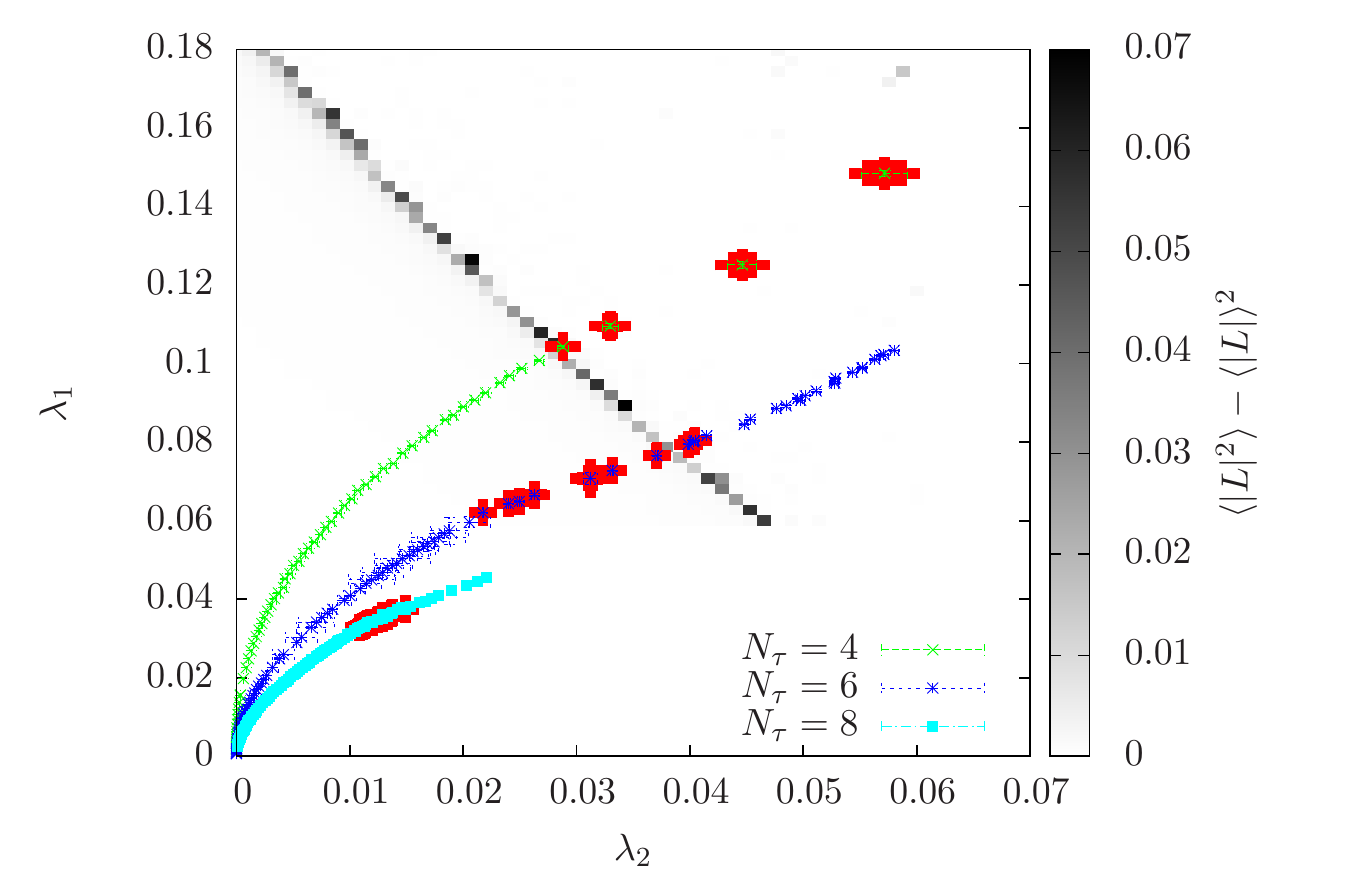}
 \end{center}
 \caption{Numerically determined $(\la_1(u(\beta)),\la_2(u(\beta)))$ for various 
 $\nt$. 
The grey line represents the phase boundary of the two-coupling effective theory.
Red points mark the critical region $0.995 u(\beta_c) <u < 1.005 u(\beta_c)$ 
in full Yang-Mills theory  for the different $\nt$.
}
 \label{fig:ntdepil}
\end{figure}

Before we turn to the more involved interaction terms, we discuss the behaviour of
 the effective couplings towards the continuum limit of the 4d Yang-Mills theory. 
Since the strong coupling series for larger distance interactions 
are parametrically suppressed by additional powers of $N_\tau$ 
compared to the nearest
neighbour interaction, it is tempting to expect their relevance to diminish with growing 
$N_\tau$ \cite{Langelage:2010yr}. However, Figure \ref{fig:longrangec} shows that at the phase transition
all couplings become of 
comparable size. This is not unexpected, since the critical coupling represents the convergence radius
of the strong coupling expansion.
The general behaviour for growing $N_\tau$ can be inferred by inspecting 
(\ref{eq:npl1l2l3t}). The $\tilde{\lambda}_i$ on the left side of the equations are given by bare Polyakov loop
correlators. For fixed temperature in the 4d Yang-Mills theory, these get smaller with 
growing $N_\tau$, i.e.~the continuum limit of  (\ref{eq:npl1l2l3t}) is zero. 
The numerical values of the couplings for $\nt=6$ are shown in Figure \ref{fig:longrangecd}. The effective couplings at a fixed $u$ are smaller than for $\nt=4$, a behaviour also shared by the corresponding 
strong coupling series. Consequently, all couplings
remain small on the way to the continuum. 

Consider now the one-coupling theory for
growing $N_\tau$. Because $\lambda_1$ is diminishing with $N_\tau$ for fixed physical $T$,
equation (\ref{eq:npl1}) will at some point no longer have a solution corresponding to the critical value
$(\lambda_1)_c$ for the phase transition in the one-coupling theory, hence a second coupling has to be
added. The two-coupling effective theory has a critical line as shown in Figure \ref{fig:ntdepil}. 
The intersection of the curves $\lambda_1(\lambda_2)$ for given $N_\tau$ now give a prediction for
$\beta_c$ in implicit form, which allows $(\lambda_1)_c$ to shrink with $N_\tau$, but at the same time 
$(\lambda_2)_c$ has to increase. This is remedied by adding a third coupling, etc. 
The general situation is thus: any effective theory with a finite number of couplings features a critical
hypersurface representing the phase boundary between an ordered and a disordered phase. 
For fixed $N_\tau$, the effective couplings $\lambda_i$ intersect these hypersurfaces to provide the predictions for the critical couplings. 
Towards the continuum limit all intersection points have to move towards
the origin, which enforces the addition of more couplings as $N_\tau$ is increased.  

\begin{figure}[t]
 \begin{center}
\subfigure[\label{fig:ntdep1a} $\nt=6$, $12^3$ lattice]{\includegraphics[width=7cm]{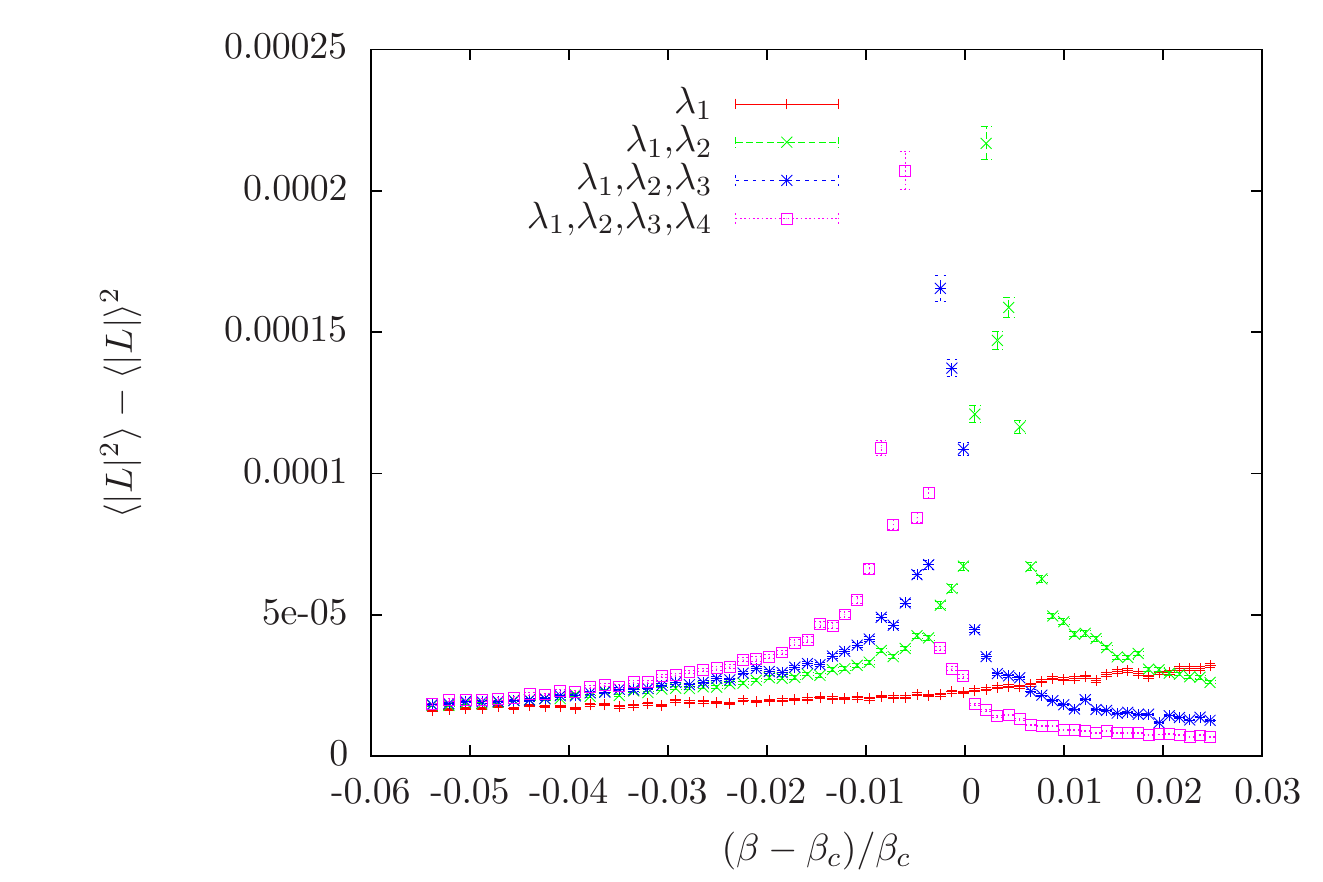}}
\subfigure[\label{fig:ntdep1b} $\nt=8$, $16^3$ lattice]{\includegraphics[width=7cm]{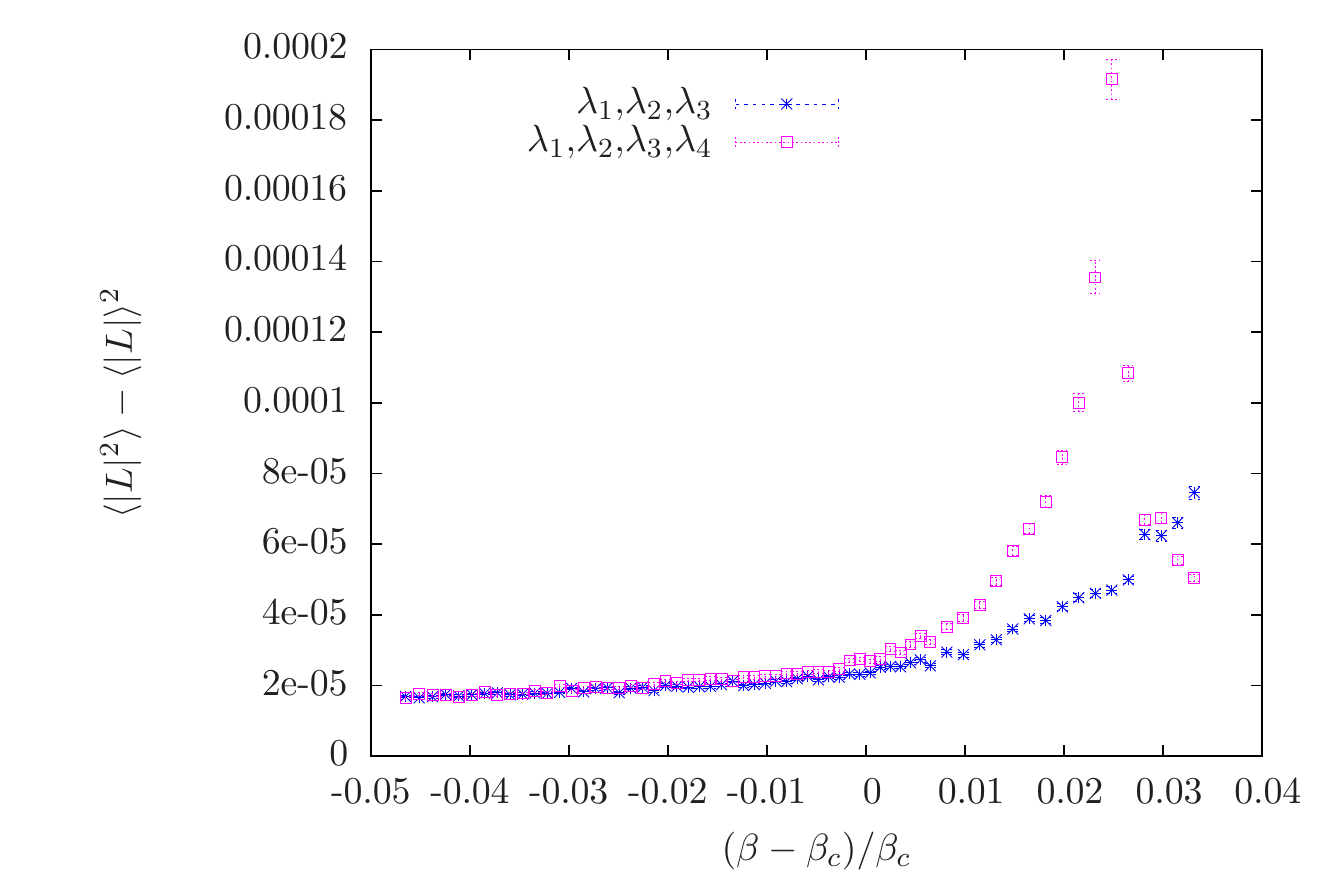}}
\end{center}
\caption{Susceptibility of the Polyakov loop. The x-axis gives the relative deviation from
($\beta_c$) of the full Yang-Mills theory. 
}
\label{fig:ntdep1}
\end{figure}
How many couplings are required in practice depends on the demanded accuracy. 
In Figure \ref{fig:ntdep1a} the critical coupling for $N_\tau=6$ with 
different truncations of the effective action is
indicated by the peak of the susceptibility. The transition point with two and more effective couplings is 
within 1\% of the full theory, while the truncation to one coupling does not show any transition in the considered range.
The importance of interaction terms at larger distances further increases at $\nt=8$, 
Figure \ref{fig:ntdep1b}. Agreement to 1\% between effective and full theory is obtained by including the couplings up to $\la_4$. This is also consistent with Figure \ref{fig:ntdepil}. The effective couplings obtained in the 5\% critical region of Yang-Mills theory nicely coincide with the transition line in the two coupling model for $\nt=4$ and $\nt=6$. For $\nt=8$ the truncation to the two-coupling effective action is not enough for an accurate description of the phase transition.

\begin{figure}[t]
 \begin{center}
\includegraphics[width=0.49\textwidth]{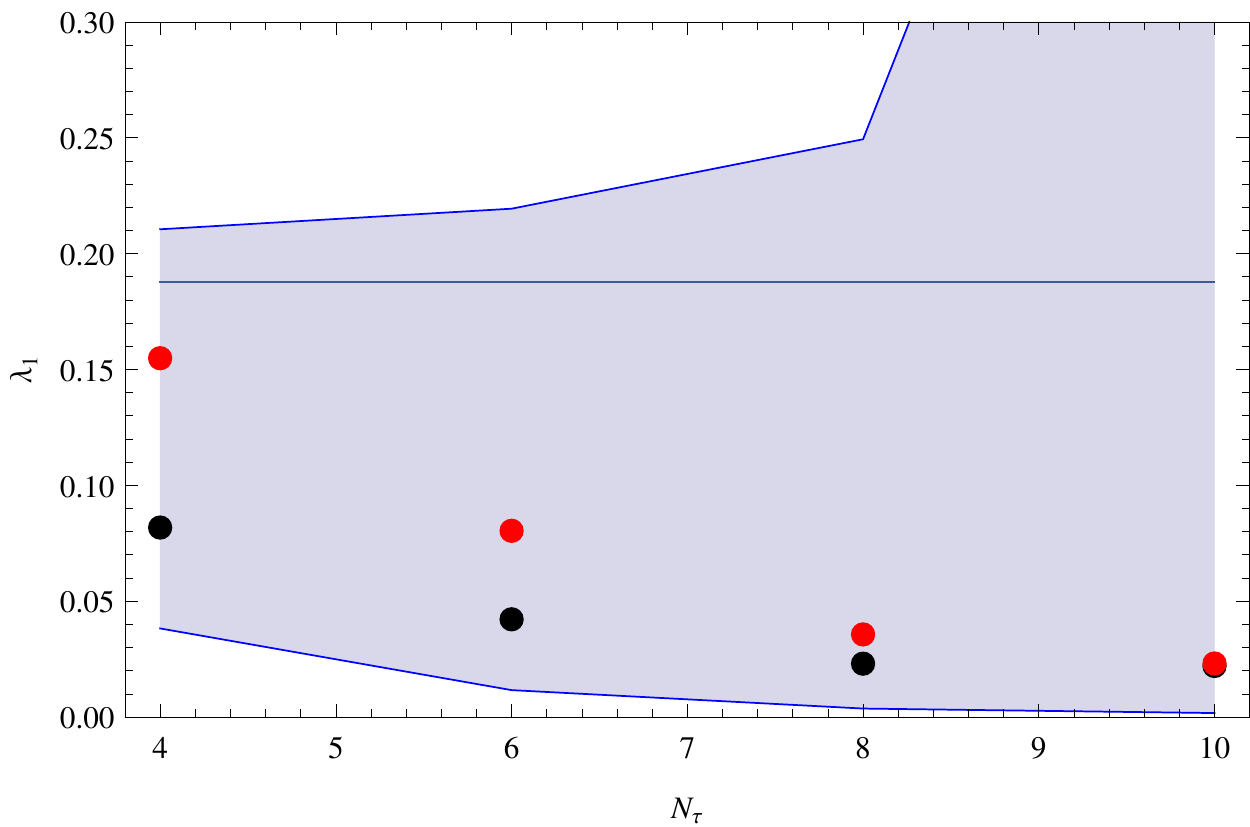}
\includegraphics[width=0.49\textwidth]{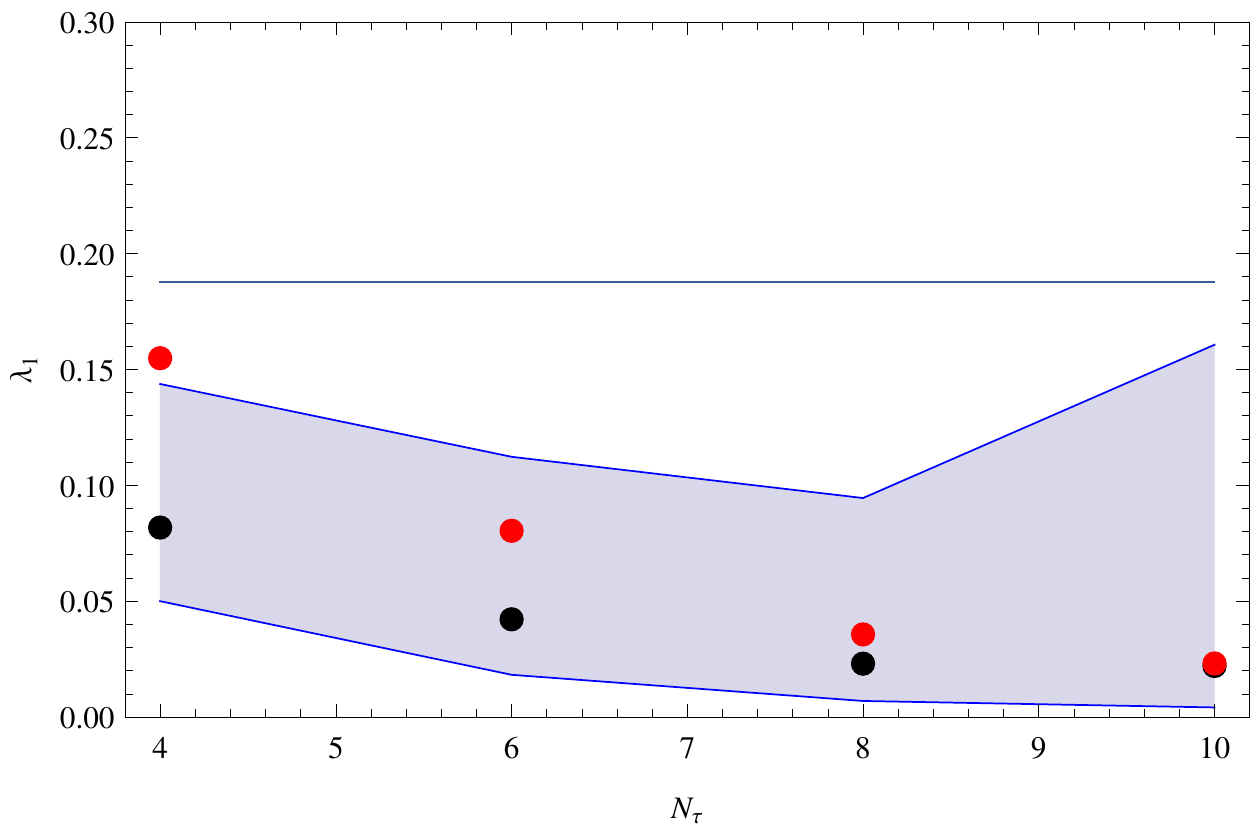}
 \end{center}
 \caption[]{Red points give the numerical solution and black points the strong coupling 
 series for $\lambda_1(\beta_c(\nt))$ . The shaded areas give the region of $\beta$ 
 within 8\% (left) or 5\% (right) of the true critical value $\beta_c$.}
 \label{fig:scband}
\end{figure}

Finally we use the numerical coupling $\lambda_1$ from our least truncated effective theory to 
assess the quality of the one-coupling theory with the analytic expression for $\lambda_1$.
Figure \ref{fig:scband} compares the numerical and analytical mappings, 
$\lambda_1(\beta_c(N_\tau))$, both evaluated at 
the true $\beta_c(N_\tau)$ of the full theory. Interestingly, the quality of the analytic function appears
to initially improve with $N_\tau$, until it breaks down for $N_\tau\geq 10$, where $\lambda_1(\beta_c)$
ceases to get smaller. Also shown are regions $\lambda_1(\beta_c\pm\Delta)$, 
where $\Delta$ amounts to 8\% (left) and 5\% (right) deviation from the true value.
The horizontal line is the critical value $(\lambda_1)_c$ of the one-coupling theory.
Hence, permitting 8\% relative error in the prediction for $\beta_c$, the one-coupling theory with the analytic expression 
for $\lambda_1$ may be used\footnote{Indeed, the predictions for $\beta_c$ from the one-coupling theory
in \cite{Langelage:2010yr} are all within 7-8\% of the true result.}, whereas more couplings and longer series are required for more
stringent accuracy requirements. 

\subsection{Functional form of the nearest neighbour coupling}

\begin{figure}[t]
\includegraphics[width=0.5\textwidth]{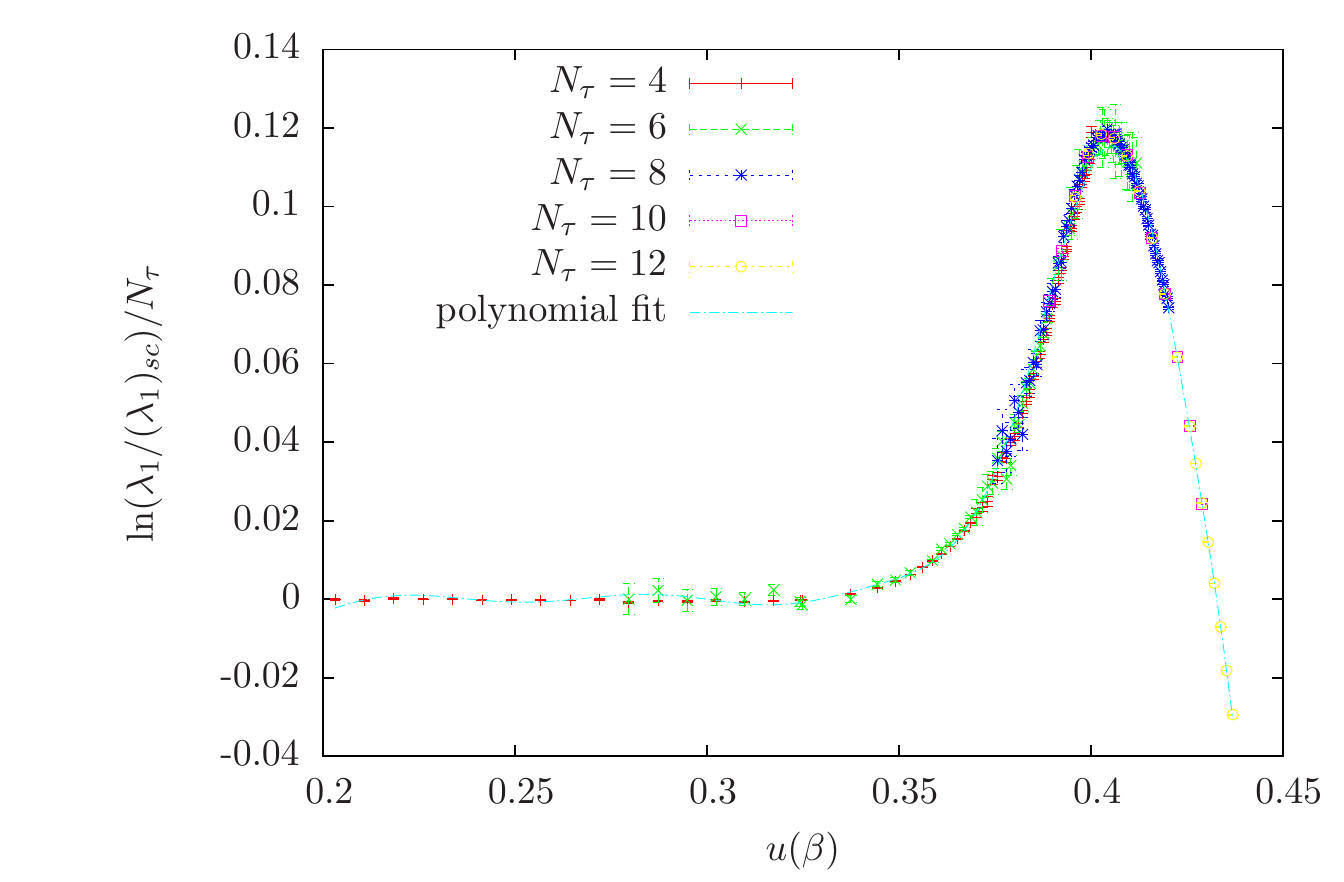}
\includegraphics[width=0.5\textwidth]{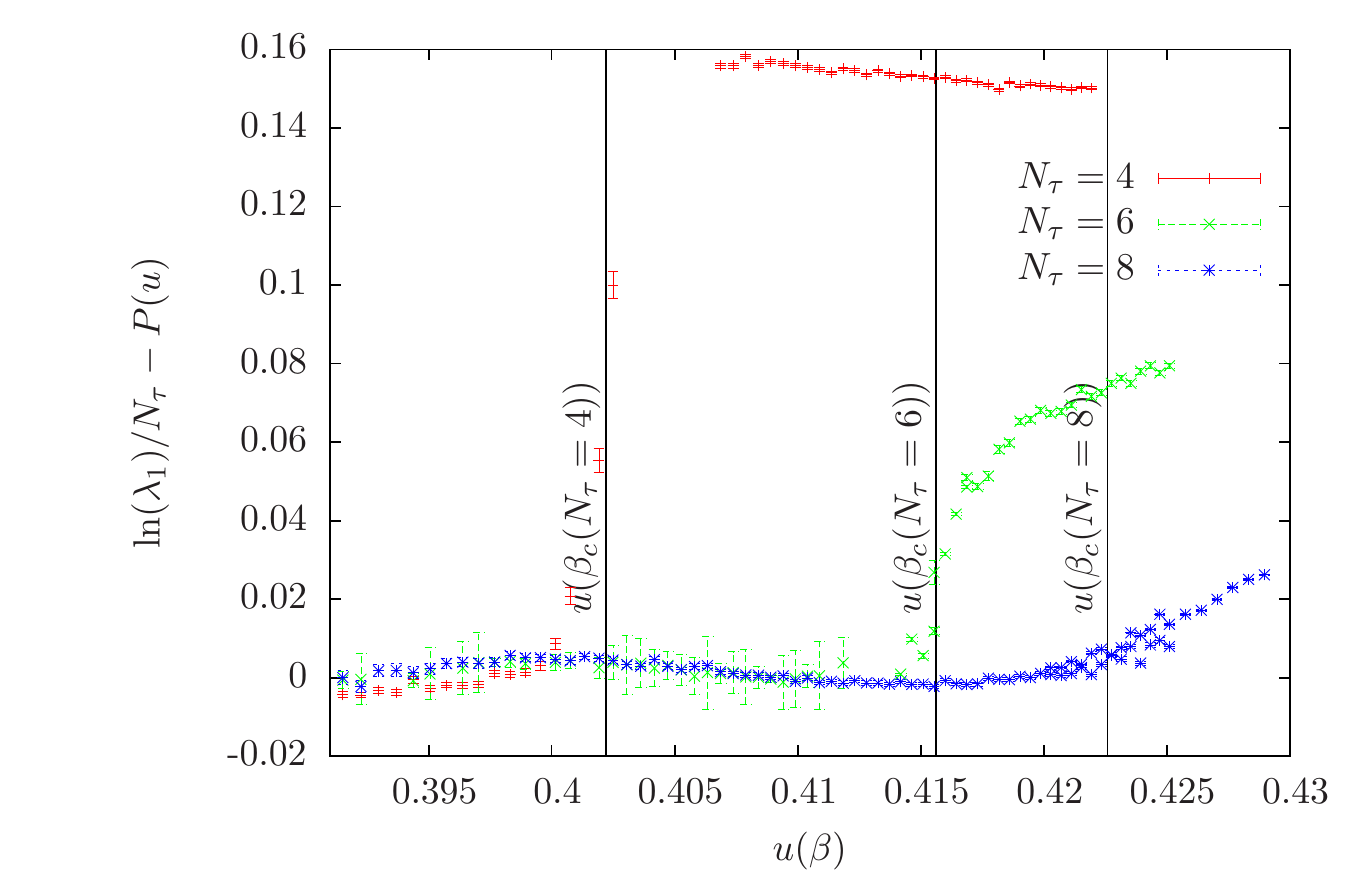}
 \caption{Left: Correction $P_c(u)$, \eqref{eq:pc}, to the strong coupling 
 result $P_{sc}(u)$ of the parametrisation \eqref{eq:l1scaling} in a logarithmic representation.
 For each $\nt$ the curves are restricted to $u<0.995 u(\beta_c)$.
 The curves contain the strong coupling results to order $u^{10}$ and 
 fits of the orders $u^{11}$ to $u^{24}$.
 Right: Deviation from the fitted polynomials in the critical region.
 The simulations were done on $4\times24^3$, $6\times24^3$, $8\times24^3$, $10\times 30^3$ , $12\times 36^3$, and, to check the finite volume effects, $6\times36^3$, $8\times48^3$.
}
 \label{fig:ntdep}
\end{figure}
After resummation of the strong coupling series, the  nearest neighbour coupling
in equation  (\ref{eq:l1sc}) is parametrised as 
\begin{equation}
 \la_1(u,\nt)=u^{\nt}\exp\left(\nt P(u,\nt)\right)\; ,
 \label{eq:l1scaling}
\end{equation}
with $P(u,\nt)$ a polynomial in $u$ for every value of $\nt$.
Between $\nt=4$ and $\nt=6$ the coefficients of these polynomials differ only at order $u^{10}$. 
For larger $\nt$ the difference is at even higher order  
such that $P(u,\nt)\approx P(u)$ can be considered as approximately independent of $\nt$,
as long as the expansion and resummation converge. 

In order to check the validity of the parametrisation \eqref{eq:l1scaling}, consider the coupling $\lat_1$ corresponding to the Polyakov line correlator, which is related to the 
renormalised free energy of a static quark anti-quark pair at a distance $R$ of one lattice spacing $a$,
\begin{equation}
 \lat_1=\erw{L((0,0,0))L((1,0,0))^\dag}=Z(u)^{\nt}\exp(-F_R(R=a,T)/T)\; ,
  \label{eq:p}
\end{equation}
with $Z(u)$ the multiplicative renormalisation factor.
To leading order $\la_1=\lat_1$, which implies
\begin{equation}
  P(u)\approx \ln(Z(u)/u)-a F_R(R=a,T)\; .
\end{equation}
The parametrisation \eqref{eq:l1scaling} with a $\nt$-independent polynomial is hence appropriate as long as $aF_R(R\nolinebreak=\nolinebreak a,T)$ does not show a strong $\nt$ dependence, which 
is generally true for the Coulomb part of the short distance region, $RT\ll 1$.

To investigate the validity of the parametrisation quantitatively, we calculate the correction $P_c$ to the polynomial $P_{sc}$ of the strong coupling approach, 
\begin{equation}
P(u)=P_{sc}(u)+P_c(u)\;,
\label{eq:pc}
\end{equation}
from the numerically determined $P(u)$.
$P_c(u)$ is shown in Figure \ref{fig:ntdep} (left) based on simulations with $\nt=4$, $6$, $8$,
$10$, and $12$ for  $u<0.995 u(\beta_c(\nt))$. 
In this entire range the data for $P(u)$ are indeed 
$\nt$-independent to a very good approximation. 
In the immediate vicinity of the phase transition the parametrisation \eqref{eq:l1scaling} breaks down,
as the zoom into this region in Figure  \ref{fig:ntdep} (right) shows.  
This $\nt$-dependence is associated with the $T$ dependence of $aF_R(R=a,T)$, which in the 
transition region gets more pronounced due to the increasing influence of the deconfined phase.
The strong coupling series has only information from the confined phase and hence underestimates 
this effect near the radius of convergence.

Figure \ref{fig:ntdep} (right) illustrates that the $\nt$-dependence of $P(u)$ is reduced with 
increasing $\nt$. 
This also follows from \eqref{eq:p}: the larger $\nt$, the smaller is the distance 
$R=a$ in physical units and the short range part of the free energy is dominated by the 
$T$-independent Coulomb contributions. At the same time the interactions at larger distances become increasingly relevant at the transition point.

The parametrisation by $P(u)$ serves to extract the predictions for the critical coupling $\beta_c$ 
from the one-coupling effective theory as in \cite{Langelage:2010yr}. 
As long as its highest order coefficient is positive, there is always a $\beta_c$ for which
\begin{equation}
 \sqrt[\nt]{\la_1(u(\beta_c),\nt)}=u\exp(P(u))=\sqrt[\nt]{(\la_1)_c}\;.
\label{eq:matchlc}
\end{equation}
Conversely, \eqref{eq:matchlc} with the fitted polynomial $P(u)$ can also be used to 
approximately compute 
$\la_1(u(\beta),\nt)$, assuming that the one-coupling effective
 action is enough for a reasonable approximation of the full theory. 
\begin{figure}[t]
 \begin{center}
\includegraphics[width=12cm]{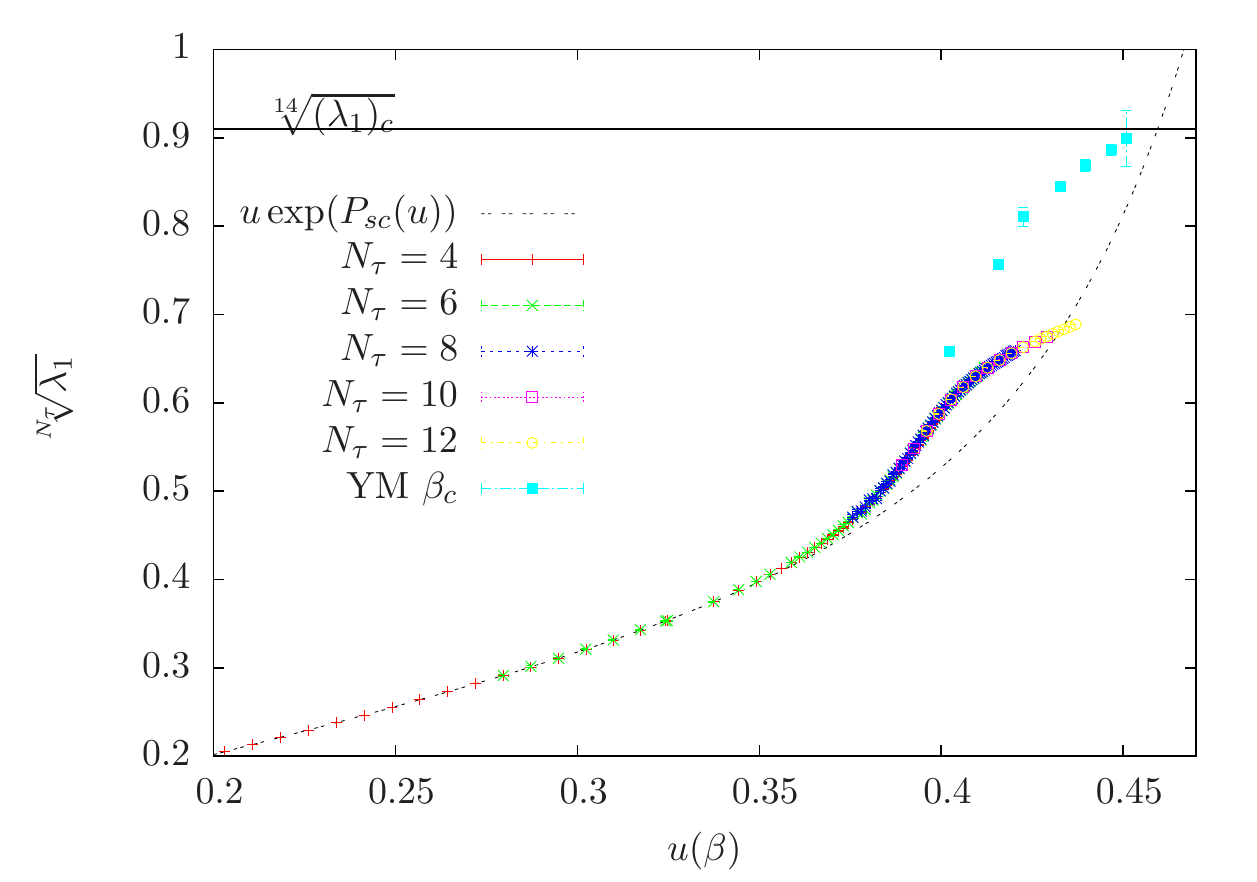}
 \end{center}
 \caption{Same data for $\lambda_1$ as Figure \ref{fig:ntdep}. 
In addition, the effective coupling determined by a matching of the 
transitions in the effective and full theory is shown (YM $\beta_c$). 
These values are a solution of \eqref{eq:matchlc} for $\lambda_1$ with $\beta_c$ 
determined in simulations of the full theory \cite{bcrit1,bcrit2,bcrit3}.
The leading $\nt$ dependence has been removed assuming \eqref{eq:l1scaling}.
The horizontal line corresponds to the transition in the effective one-coupling theory for $\nt=14$.}
 \label{fig:ntdepex}
\end{figure}

Using the parametrisation \eqref{eq:l1scaling}, the $\nt$-dependence can be factored out and the 
predictions for $\la_1$ at different $\nt$ can be combined as in Figure \ref{fig:ntdepex}. Close to the 
phase transition the numerically evaluated effective coupling is smaller than the values 
suggested by the matching of the phase transitions of the full and the effective one-coupling 
theory using \eqref{eq:matchlc}. This deviation is due to the neglected $\nt$-dependence of the 
parametrisation \eqref{eq:l1scaling} close to the phase transition, cf.~Figure \ref{fig:ntdep} (right), as well
as the increasing relevance of the interactions at larger distances. The analytic strong coupling 
prediction shows the correct dependence of $\la_1$ on $u$ and $\nt$ for small $u$. 
At larger $u$ it first underestimates and then overestimates the slope of the numerical curve. Due to this fact there is an intersection point between the expected transition line from the matching of the full and effective theory and,
around $\nt=14$, the one-coupling action with effective coupling of the strong coupling approach reproduces the critical coupling of the full Yang-Mills theory.  In the region of $\nt=8$ to $\nt=14$ the full and the effective theory data seem to converge. 

By contrast, the numerically determined effective coupling suggests that higher order coefficients of $P(u)$ are negative, as for $P_{sc}$ truncated at order $u^{7}$. Hence a simple continuation of $P(u)$ does not provide an intersection point with condition \eqref{eq:matchlc}
and the effective theory truncated to the nearest neighbour interaction fails to reproduce the transition of the underlying full theory at larger $\nt$, as discussed in the previous section.

\section{Higher representations and n-point interactions}
\label{sec:hirep}

An effective action in terms of two-point interactions seems to provide reasonable results for the discussed correlation functions and the phase transition for $\nt=4$
and $6$. On the other hand, when interactions over two and more lattice spacings are included, the restriction to two-point functions is no longer justified within in the strong coupling expansion. 
Terms involving higher $n$-point couplings appear at the same order as the long range interactions. 
In the following we
investigate the contributions of higher representations 
as well as $n$-point interaction terms.

\subsection{Strong coupling results}

The $n$-point interaction with $n>2$ and higher representations of the Polyakov loop are 
parametrically suppressed 
compared to the two-point nearest neighbour fundamental interaction.
However, relevant contributions appear already at the same order as the coupling
$\lambda_2$, such as the two-point interactions of higher representations,
\begin{align}
 \la_{(1,1)}\equiv\la^{(1,1)}_{[(1,0,0),(1,1)]} &= v^\nt(1+ \frac{9}{2}\nt \frac{u^6}{v})\label{eq:adj}\;, \\
 \la_{(2,0)}\equiv\la^{(2,0)}_{[(1,0,0),(0,2)]} &= w^{\nt}\left(1 + 6\nt \frac{u^6}{w}\right)\label{eq:sext}\; ,\\
 \text{ where }& v=\frac{9}{8}(u^2-u^3)+\frac{81}{32} u^4+O(u^5)\\
  \text{ and }& w=\frac{3}{4}u^2-\frac{9}{16}u^4+\frac{297}{80} u^5+O(u^6)\; .
\end{align}
Polyakov loops of higher representations with n-ality zero can have non-vanishing expectation values in the confined phase.
These terms appear in the effective theory as one-point couplings. The simplest example is the one-point term of the adjoint loop, 
\begin{equation}
 \la^{(1,1)} = 108 \nt (\nt+1)u^{4\nt+2}\; .
\end{equation}
Since it is of high order we neglect it in our considerations. 

In addition, there are higher n-point interactions in the fundamental representation. 
The leading three point interaction is a combination of fundamental, anti-fundamental, and adjoint loop:
\begin{equation}
 \lambda^{(1,0)}_{[(1,0,0),(1,1)],[(1,1,0),(0,1)]}=-4\nt u^{2\nt+8}\; .
\end{equation}
This is beyond the order considered in our current investigations.
Four-point interactions can appear in terms of two fundamental and two anti-fundamental loops.
In the leading contribution these loops are located on the corners of a plaquette, with
two inequivalent configurations 
\begin{align}
  \la_p\equiv\la^{(1,0)}_{[(1,0,0),(0,1)],[(1,1,0),(1,0)],[(0,1,0),(0,1)]}&=\frac{1}{2}\nt(\nt-1)u^{2\nt+2}\label{eq:lap} \\
  \la^{(1,0)}_{[(1,0,0),(1,0)],[(1,1,0),(0,1)],[(0,1,0),(0,1)]}&=\nt u^{2\nt}(v^4 + w^4) \; .
\end{align}
The second contribution is beyond the maximal order considered in this work and neglected.
It is obvious at this point that there are many different configurations for the four-point interaction 
in addition to the above plaquette form. 
These appear at $u^{2\nt+n}$ for $n>4$ and are also neglected here.
Once long range interactions of the fundamental and anti-fundamental loop beyond distance three 
are included, omission of these terms is no longer justified.

The effective action with the considered additional terms has the following form
\begin{eqnarray}
 e^{-\Sefft}&=&e^{-\Seffo}\prod_{x,i=1,\ldots 3} \Big(1+\la_{(2,0)} [L_6(x) L_6(x+\hat i)^\dag+L_6(x)^\dag L_6(x+\hat i)]\Big)\\
 && \prod_{x,i=1,\ldots 3} \Big(1+\la_{(1,1)} L_8(x) L_8(x+\hat i)\Big)\\
 && \prod_{x,i=1,\ldots 3,j<i} \Big(1+\la_p [L(x) L^\dag (x+\hat j)L (x+\hat i+\hat j)L^\dag (x+\hat j)+\text{c.c.}]\Big)\; .
 \label{eq:prodadjsexplaq}
\end{eqnarray}
We have introduced a short hand notation for the sextet ($L_6(\ix)\equiv\chi_{(2,0)}(W(\ix))$) and adjoint ($L_8(\ix)\equiv\chi_{(1,1)}(W(\ix))$) Polyakov loop.

\subsection{Numerical calculation of the effective couplings}

\begin{figure}[ht]
 \begin{center}
\includegraphics[width=12cm]{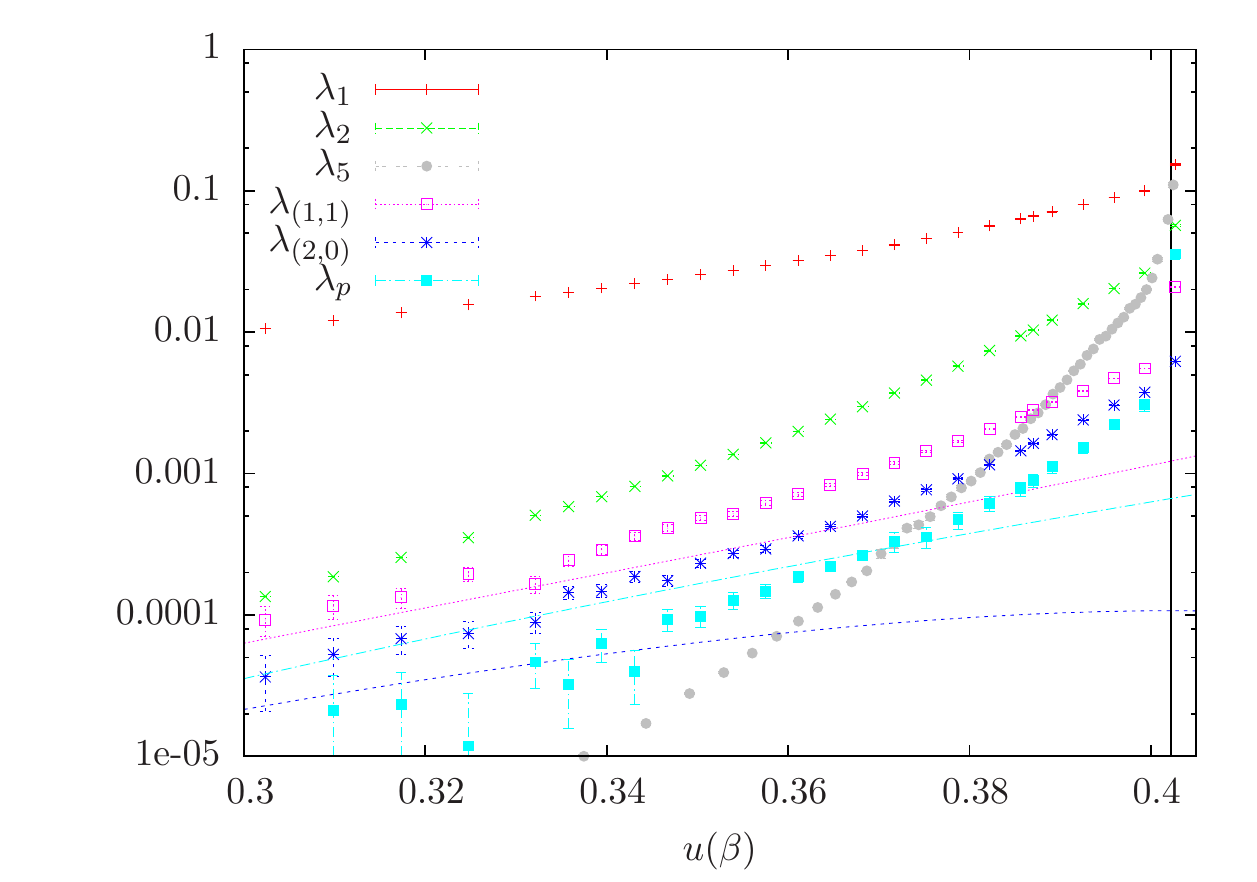}
 \end{center}
 \caption{The coupling constants of the adjoint next neighbour, sextet next neighbour, and plaquette interactions in the effective theory \eqref{eq:prodadjsexplaq}. These constants are a solution of the equations \eqref{eq:sexadj}. The lines are the strong coupling results \eqref{eq:adj}, \eqref{eq:sext}, and \eqref{eq:lap}. The vertical line indicates $\beta_c$ of full Yang-Mills theory. The simulations have been done on a $4\times24^3$ lattice. Results for $\la_1$, $\la_2$, and $\la_5$ of the fundamental loop are shown for comparison. }
\label{fig:adjsext}
\end{figure}
%%%%%%%%%%%%%%%%%%%%%%%%%%%%%%%%%%%%%%%%%%%%%%
The additional couplings are related to the correlators of Polyakov loops in higher representations and
the four-point correlator of fundamental Polyakov loops. The coefficients $\lat$ that follow directly from
these measurements are
\begin{equation}
\begin{split}
 \lat_{(1,1)}&=\erw{\chi_{(1,1)}(W((0,0,0)))\chi_{(1,1)}(W((1,0,0)))}\\
 \lat_{(2,0)}&=\erw{\chi_{(2,0)}(W((0,0,0)))\chi_{(0,2)}(W((1,0,0)))}\\
 \lat_p&=\erw{L((0,0,0))L^\dag((1,0,0))L((1,1,0))L^\dag((0,1,0))}\;.
\end{split}
\end{equation}
As in the previous case we expand these expectation values in the effective couplings $\lambda_n$. 
Along the same lines as for the couplings $\lat_1$ to $\lat_5$ additional equations can be derived 
that enlarge the system of equations \eqref{eq:npl1l2l3t}.
The correlators of the adjoint and sextet representation receive contributions from  the fundamental interactions  only at high orders.
In contrast, the plaquette correlator contains the low order two-point fundamental interaction in its disconnected part.
At the current order in our investigations we get the following set of additional equations
\begin{equation}
 \begin{split}
  \lat_{(1,1)}&=\la_{(1,1)}+8\la_1^4+16\la_1^2\la_2+\ldots\\
  \lat_{(2,0)}&=\la_{(2,0)}+4\la_1^4+8\la_1^2\la_2+\ldots\\
  \lat_p&=\la_p+2\la_1^2+12\la_1^4+24\la_1^2\la_2+\ldots\; .
 \end{split}
\label{eq:sexadj}
\end{equation}

The numerical solution for the additional couplings from simulations at $\nt=4$ is shown in Figure \ref{fig:adjsext}. 
In the region of small $\beta$ the couplings of the adjoint and sextet representation as well as the four-point 
interaction are larger than the long distance 
couplings in the fundamental representation. According to our estimates, this is not compensated 
by the larger number of neighbours for the fundamental Polyakov loop interaction at larger distances. 
On the other hand, we observe a stronger increase of the fundamental couplings 
as $\beta$ gets larger and a change of the ordering of the interactions towards the 
phase transition.

\section{Conclusions}
\label{sec:conc}

We have presented a new numerical method to determine the couplings of three-dimensional 
effective Polyakov loop lattice theories for $\text{SU}(N)$ Yang-Mills systems at finite temperature, 
which arise by integrating out the spatial degrees of freedom.
The method is based  on the use of two different character expansions.
In the first, the effective couplings are identical to definite $n$-point functions
of Polyakov loops which can be easily simulated in full Yang-Mills theory. While the corresponding 
truncated effective theory has (within statistical accuracy) exact couplings, it will 
reproduce correlation functions only up to a maximal distance, depending 
on the truncation, beyond which all $n$-point functions vanish.
%Therefore it cannot be used to describe phase transitions.
A second character expansion, commonly used in strong coupling approaches, contains
long range correlations in any truncation, but its couplings are not directly simulable. 
Both sets of effective couplings are small in the entire parameter range and related by a perturbative expansion with high accuracy,
where the ordering of couplings is guided by strong coupling power counting.
The resulting effective theory has the same structure as the one determined in a strong coupling
expansion \cite{Langelage:2010yr}, but with non-perturbatively improved couplings which
can also be determined in the deconfined phase. 
A large number of effective couplings
can be extracted by simulations with relative ease compared to an analytic strong coupling
series, but of course this has to be redone for every set of parameter values $(\beta,\nt)$. 

%The proposed method allows for a thorough test of the analytically determined effective couplings,
%as well as the determination of numerically improved ones.
We find the strong coupling series for the nearest neighbour interaction 
to agree with the improved coupling up to 
the immediate neighbourhood of the phase transition,
which marks the convergence radius of the strong coupling series. 
As already observed in \cite{Bergner:2013qaa}, a quantitative description of the Polyakov loop 
correlator requires the non-local couplings up to the lattice distance of interest. 
Including these with our new
method, quantitative agreement is achieved.   
Also near the phase transition, we observe an increasing relevance of long-range couplings, 
which can be related to the fact that the correlation length is maximal there. 
For $\nt=4$ and $\nt=6$ the relevant contributions 
are the nearest and next-to-nearest neighbour interactions, again in accord with earlier 
observations \cite{Bergner:2013qaa}. As the lattice gets finer, more interactions 
have to be included, depending on the desired accuracy. 
The critical coupling can be obtained at 10\% accuracy from the one-coupling theory with
the analytic effective coupling in a range $\nt=2-10$. 
%For $\nt>10$ or higher accuracy, additional
%couplings need to be taken into account. 
Conversely, with our numerical determination of the 
effective couplings, effective theories reproduce the location of the phase transition 
to 1\% accuracy with one coupling for $\nt=4$ and five couplings at
$\nt=8$. It would be very interesting to extend this approach to the fermionic 
sector of QCD.

\section*{Acknowledgements}
We thank Wolfgang Unger for numerous enlightening discussions.
G.B.~and O.P.~are supported by the German BMBF, No.\ 06FY7100 as well as
Helmholtz International Center for FAIR within the LOEWE program of the State of Hesse.
The authors acknowledge the computing time granted by the LOEWE-CSC of 
Goethe-University Frankfurt.

\end{document}